\documentclass[preprint]{article}
\usepackage[margin=2cm]{geometry}

\begin{document}

\title{Kaluza-Klein Higher Derivative Induced Gravity}
\author{
W. F. Kao\thanks{%
wfgore@cc.nctu.edu.tw} \\
Institute of Physics, Chiao Tung University, Hsinchu, Taiwan}
\maketitle

\begin{abstract}
The existence and stability analysis of an inflationary solution
in a $D+4$-dimensional anisotropic induced gravity is presented in
this paper. Nontrivial conditions in the field equations are shown
to be compatible with a cosmological model in which the
4-dimension external space evolves inflationary, while, the
D-dimension internal one is static. In particular, only two
additional constraints on the coupling constants are derived from
the abundant field equations and perturbation equations. In
addition, a compact formula for the non-redundant $4+D$
dimensional Friedmann equation is also derived for convenience.
Possible implications are also discussed in this paper.
\end{abstract}


\section{Introduction}
Higher-order corrections to the Einstein gravity \cite{kim,dm95}
can be derived from the quantum gravity and the string theory
\cite{green}. Application to the study of the inflationary
universe \cite{acc,acc1} have been a focus of research interest.
In particular, higher derivative terms also arise as an effective
theory for the quantum corrections of matter fields in a curved
space\cite{green}.

In addition, Kaluza-Klein theory \cite{kk,visser} is also
important in the study of the evolution of the early universe.
Indeed, the dimensional-reduction process could affect the
evolution of inflation universe significantly. Recently, brane
universe scenario has also become a focus of interest \cite{ib}.

Induced gravity models have been a focus of study for many
reasons. Weyl is the first to propose that the scale invariant
theory is a candidate for the unified theory of gauge field and
gravitational field. In addition, Dirac's large number theory also
asserts that all dimensionful parameters in physical theory are in
fact dynamical functions of time. As a result, various interesting
models have attracted researchers' interest for a long time.

Therefore, we intent to study an $N=4+D$ dimensional Kaluza-Klein
higher derivative induced gravity model with all dimensionful
coupling constants replaced by appropriate scale-dependent fields.
We will show that a constant internal-space solution will lead to
a nontrivial constraint to the field equations. In addition, there
is also another constraint derived from the assumption of constant
internal scale-dependent scalar field $\psi$.

In fact, we will show in this paper that there are three
constraints to be imposed on the choice of three different
coupling constants coupled to the higher curvature terms. Note
that, in four-dimensional space, $ E =
R_{\lambda\mu\nu\rho}\,R^{\lambda\mu\nu\rho} -
4\,R_{\mu\nu}\,R^{\mu\nu}+ R^2 $ is the integrand of the
Gauss-Bonnet term \cite{weinberg1}. In addition, Weyl tensor
\cite{weinberg1} connects these fourth-order curvature terms in
four-dimensional space. Hence we only need to deal with $R^2$ term
in four-dimensional de Sitter space. These constraints does not
hold in higher dimensional spaces. Therefore, we must deal with
all three different fourth-order terms in $N$-dimension. Hence
equations of motion for the higher-derivative Kaluza-Klein induced
gravity theory\cite{zee,ni,kao00,smolin} are much more complicated
than the four-dimensional higher-derivative gravity. We will show,
however, that the mentioned abundant constraints are not only
consistent with the already abundant stability constraints of the
system but also lead to an interesting result: i.e. only the $R^2$
coupling term is consistent with the inflationary solution.
Possible implications will also be discussed in this paper.

In order to reduce the complication of the derivation of field
equations, we will also derive a simple expression for the
Friedmann equation \cite{kao99} in a ($4+D$)-dimensional space in
this paper. The redundancy of the associated field equations due
to the Bianchi Identity will also be analyzed. We will show that
quadratic terms do not affect the Friedmann equation in a constant
flat internal space scale factor $d(t)$ and flat de Sitter four
space.

This paper will be organized as follows: (i) In section II, we
will introduce the Kaluza-Klein higher derivative induced gravity
model with all dimensionful parameters replaced by appropriate
dynamical fields. The constraint derived from the constant $\psi$
field will be obtained in this section. (ii) We will derive a
model-independent expression for the $N$-dimensional generalized
Friedmann equation in the higher-dimensional higher-derivative
theory in section III. These formulae are derived from a reduced
one-dimensional theory. (iii) We will discuss the stability
conditions for an inflationary solution in the induced-gravity
theory in section IV. (iv) Finally, conclusions are presented in
section V.

\section{Kaluza-Klein Higher Derivative Induced Gravity}
The $4+D$-dimensional Kaluza-Klein higher derivative induced
gravity theory can be described by the following Lagrangian:
\begin{equation} \label{Lind}
L_N =\left ( L- {1 \over 2} \partial_A \psi \partial^A \psi \right
) \psi^D.
\end{equation}
with the $4$-dimensional relevant higher derivative induced
gravity Lagrangian $L$ given by:
\begin{equation}
L=L_1+ L_2+L_\phi=- {\epsilon \over 2}  \phi^2 {\bf R}   -c_1
({\bf R}^{AB}_{\;\;\;\;\;\;CE})^2 - c_2 ({\bf R}^{A}_{\;\;\;B})^2
- c_3 {\bf R}^2- {1 \over 2} \partial_A \phi \partial^A \phi
-V(\phi) \label{pdg}
\end{equation}
Here $L_1 \equiv - \epsilon  \phi^2 {\bf R} /2$, $L_2 \equiv -c_1
({\bf R}^{AB}_{\;\;\;\;\;\;CE})^2 - c_2 ({\bf R}^{A}_{\;\;\;B})^2
- c_3 {\bf R}^2$, and $L_\phi \equiv - (\partial \phi)^2/2
-V(\phi)$ denote the induced Einstein-Hilbert Lagrangian, higher
derivative terms, and the scalar field Lagrangian.

Note that, throughout this paper, the curvature tensor ${\bf
R}^E_{ABC} ({\bf g}_{AB})$ will be defined by the following
equation
\begin{equation}
[D_A,\,\, D_B]V_C = {\bf R}^{E}\,\,_{CBA} V_E . \label{eqn:curv}
\end{equation}
Accordingly, ${\bf R}^{E}_{ABC} = - \partial_C {\bf \Gamma}
^E_{AB} - {\bf \Gamma} ^F_{AB} {\bf \Gamma} ^E_{CF} - ( B
\leftrightarrow C)$. Here ${\bf \Gamma} ^C_{AB}$ denotes the
Christoffel symbol (or spin connection of the covariant
derivative, i.e. $D_A V_B \equiv \partial_A V_B - {\bf \Gamma}
^C_{AB}V_C$). To be more specific, ${\bf \Gamma}^C_{AB} = {1 \over
2} {\bf g}^{CE} (\partial_A {\bf g}_{EB} + \partial_B {\bf g}_{EA}
- \partial_E {\bf g}_{AB} )$. Moreover, the Ricci tensor ${\bf
R}_{AB}$ is defined as
\begin{equation}
{\bf R}_{AB} = {\bf R}^{C}\,\,_{ABC}. \label{ricci}
\end{equation}
And the scalar curvature ${\bf R}$ is defined as the trace of the
Ricci tensor ${\bf R} \equiv {\bf g}^{AB} {\bf R}_{AB}$. Also, the
Einstein tensor is defined as ${\bf G}_{AB} \equiv {1 \over 2}
{\bf g}_{AB} {\bf R} - {\bf R}_{AB}$.

Note that we will use bold-faced notation (e.g. ${\bf R}$) to
denote field variables in $N(=4+D)$-dimensional space. In
addition, normal notations will denote field variables evaluated
in the $4$ or $D$-dimensional spaces as the dimensional reduction
process $M^N \to M^4 \times M^D$ takes places. Here $M^4$ is the
four dimensional Friedmann-Robertson-Walker (FRW) space and $M^D$
is the internal space. We will assume that $M^D$ is the $D$
dimensional FRW space for simplicity. Note that the metric of the
$4$+$D$-dimensional FRW anisotropic space $M^N$ is given by
\begin{eqnarray}
ds^2 &\equiv &{\bf g}_{AB} dZ^A dZ^B \equiv g_{\mu \nu} dx^\mu
dx^\nu +
f_{mn}dz^mdz^n \\
 & =&  -dt^2
+ {a^2}(t) \Bigl( {dr^2  \over 1 - k_1 r^2} + r^2 d^3\Omega  \Bigr
) + {d^2}(t) \Bigl( {dz^2  \over 1 - k_2 z^2} + z^2 d^D\Omega
\Bigr ) . \label{eqn:frw} \label{FRW}
\end{eqnarray}
Here $ d^p \Omega $ is the solid angle $d^p \Omega \, = \,
d{\theta_1}^2 + {\sin}^2 \theta_1 \, d{\theta_2}^2 + \cdots
+\sin^2 \theta_1 \sin^2 \theta_2 \cdots \sin^2 \theta_{p-3}
d\theta_{p-2}^2$ and $k_1,\,\, k_2 \, = \, 0, \pm 1$ stand for a
flat, closed or open universe respectively. Note that we will also
write $g_{ij}=a^2 h_{ij}$ and $g_{mn} = d^2 h_{mn}$ for
convenience. Note that $\theta_i$ is the phase angle of the
$D$-dimensional spherical coordinate. For example, we have
\begin{equation}
z_1=z \sin \theta_1 \sin \theta_2 \cdots  \sin \theta_{D-2}.
\end{equation}
Note that we will write $N$-dimensional space-time coordinate as
$Z^{A} \to (x^\mu, \,\, z^m)$ with $A (= 0, 1, \cdots , N-1)$,
$\mu (= 0,1,2,3)$, and $m(=1,2, \cdots, D)$ denoting the $N$, $4$,
and $D$-dimensional space-time indices respectively. Specifically,
capital Roman letters $A,B,C, \cdots$ will denote $N$ dimensional
indices. In addition, Greek letters will denote $4$-indices while
the second half of the Roman letters will denote $D$-dimensional
space-time indices. Here we have assumed that the internal space
($z$) is independent of the external space ($x$). The only
$t$-dependence of the internal space is through the scale factor
$d(t)$.

Induced gravity proposes that all dimensionful parameters are
dynamical variables. Therefore all coupling constants in this
models, $\epsilon$ and $c_i$, are dimensionless. Indeed, the
action
\begin{equation}
\int d^4x d^Dz \sqrt{g} L_N
\end{equation}
is invariant under the global scale transformation:
$g_{AB}'=\Lambda^{-2} g_{AB}$, $\phi'= \Lambda \phi$ ,$\psi'=
\Lambda \psi$ in the absence of the potential $V$ unless $V(\phi)
\sim \phi^4$. Here $\Lambda=$ constant denotes the global scale
transformation parameter. For a practical application, one needs
to introduce a symmetry breaking potential so that a physical
scale can be picked up dynamically. As a result a physical
inflationary solution can be managed. In addition, it is easy to
observe from the scale transformation that $\psi$ is introduced to
compensate the transformation properties of the internal
$D$-space. In deed, $\psi^D d^Dz$ is made dimensionless by
construction.

The variational equation of the $\psi$ field gives
\begin{equation} \label{psieq}
D\psi^{D-1} \left ( L- {1 \over 2} \partial_A \psi
\partial^A \psi \right )+ D_A ( \psi^D \partial^A \psi) =0.
\end{equation}
After the dimensional reduction process takes place, the $\psi$
field is expected to be a function of internal space coordinate
$z$ only.  Consequently, this internal space scalar field will not
affect the $4$-dimensional physical universe thereafter. For
simplicity, one will assume that a constant solution
$\psi=\psi_0=$ constant is adopted so that the constant $\psi$
field can be absorbed into the internal coordinate $z$ by a proper
re-scaling. In effect, the $\psi=$ constant solution introduces a
physical scale of the internal space. Note that the $\psi=$
constant is not only a solution to Eq. (\ref{psieq}), it is also a
solution consistent with the static internal space solution.
Indeed, the scalar field $\psi$ is responsible for the internal
space dimension such that $dz^D \psi^D$ remain dimensionless. As a
comparison, the scalar field $\phi$ is introduced to take care of
the dynamical dimension of all $4$-space coupling constants. As a
result, the $\psi$ field will decouple from the $4$-space
completely after the
 dimensional reduction process is completed. The remaining impact of
this solution is an additional constraint $L=0$ to be made
compatible with the dimensionally-reduced $4$-space of interest.

Therefore, we will focus on the $4+D$-dimensional model given by
the effective Lagrangian density $L$ described by Eq. (\ref{pdg}).
Consequently, we will study the model (\ref{pdg}) in the presence
of the constraint $L=0$ to be imposed later. In fact, we will show
that the constraint $L=0$ is consistent with the inflationary de
Sitter solution binding by abundant and interesting constraints to
be imposed on the coupling constants $c_i$. We will also study the
stability of this inflationary solution and discuss interesting
implications of this Kaluza-Klein induced gravity model.

The Euler-Lagrange equation of the system can be obtained from the
variational equation of the $4+D$-dimensional metric ${\bf
g}_{AB}$. We will write it as
\begin{equation}
{\bf J}_{AB}={\bf G}_{AB}-{\bf T}_{AB}=0.
\end{equation}
The derivation is very complicate and delicate. Fortunately, if we
are only interested in the $4$+$D$-dimensional FRW space, the
dynamical variables reduce to a  set of one-dimensional variables
$a(t)$ and $d(t)$ and $\phi(t)$.

Effectively, we can write the Lagrangian $L(\;{\bf g}_{AB}(a(t),
d(t)), \phi(t)\;) \to L_r(\; a(t), d(t), \phi(t)\;)$. Hopefully,
the final expression of the Euler-Lagrange equations can be
reproduced from the variation of $L_r$ with respect to the
dynamical variables $a(t)$ and $d(t)$ and $\phi(t)$.

If this is indeed applicable, the field equations can be derived
more easily without involving complicate tensor algebra. In
particular, this will be more easy to access when complicate
interactions are introduced. We will show that there is a little
problem with this approach for the non-redundant Friedmann
equation. Fortunately, the non-redundant Friedmann equation can be
reconstructed by restoring the ${\bf g}_{tt}$ variable. In a
moment, we will derive a set of modified formulae for this reduced
Lagrangian $L_r$. For simplicity, we will drop the subscript $_r$
in $L_r$ for simplicity and economics of notations.

Note also that the Bianchi identity $D_MG^{MN}=0$ and the
energy-momentum conservation law $D_MT^{MN}=0$ implies that
$D_MJ^{MN}=0$. In addition, the $tt$ component of this Bianchi
identity can be brought to the following form:
\begin{equation}
(\partial_t + 3 H +DI) J_{tt} + 3 a^2H J_{3} +D d^2 I J_D=0,
\label{h3}
\end{equation}
with $J_{ij} = J_3 h_{ij}$. Since we only need two field equations
for the one-D system of $a(t)$ and $d(t)$. Therefore, one of the
three field equations $J_{tt}=0$, $J_3=0$ and $J_D=0$ is
presumably redundant. They are, however, not equally redundant.
Let us first assume that $H \ne 0$ $I \equiv \dot{d}/d \ne 0$ for
simplicity. Indeed, the Bianchi identity implies that the first
equation $J_{tt}=0$ is truly non-redundant: (i) $J_{tt}=0$ implies
that $3 a^2H J_{3} +D d^2 I J_D=0$ for all times. Hence the
constraint $J_3=0$ (or $ J_D=0$) implies the vanishing of the
other equation $J_D=0$ (or $J_3=0$). (ii) $J_3=J_D=0$ implies that
$ (d/dt +3H +DI) J_{tt} =0$. Hence we have $J_{tt}=$ constant
$\exp[-a^3 d^D]$ which does not go to zero unless $a^3d^D \to
\infty$.  Cases (i) and (ii) mean that the Friedmann equation
$J_{tt}=0$ is truly non-redundant while $J_3=0$ and $J_D=0$ are
equally redundant. Therefore, we can ignore either one of the
equations without losing any information. This is, however, not
true under the condition where $d=$constant, or $I \equiv
\dot{d}/d=0$. As indicated by Eq. (\ref{h3}), we have instead
\begin{equation}
(d/dt +3H) \bar{J}_{tt} = 3 a^2H \bar{J}_3 \label{bi3}
\end{equation}
under this condition. Here we have written $\bar{J}_{tt} \equiv
 J_{tt}|_{I=0}$ and similarly ${\bar J}_3\equiv
 J_{3}|_{I=0}$.  Therefore, the reduced Bianchi identity
(\ref{bi3}) only tells us that $\bar{J}_3=0$ is redundant as
compared to the non-redundant Friedman equation $\bar{J}_{tt}=0$
under the constraint $I=0$. In fact, the $d$-equation or the
$a$-equation has to be retained for a consistent check in order to
make sure whether the system does accommodate a constant $d$
solution. This point is often overlooked and should be checked
carefully in the analysis of the Kaluza-Klein theory under a
constant internal-space solution.

Bianchi identity implies that the $N(N+1)/2$ Einstein equations
$G_{AB}=T_{AB}$ are not all independent, but related by $N$
constraint equations $D_AG^{AB}=0$. By looking at the conservation
law of energy momentum tensor $D_AT^{AB}=0$, one may interpret
that conservation of energy momentum tensor implies the vanishing
of $D_AG^{AB}$. The Bianchi identity has, however, a more
intrinsic geometric meaning. It is in fact a geometric
conservation law. Note that the Bianchi identity $D_AG^{AB}=0$ has
a simple geometric interpretation, namely, the boundary of a
boundary is zero \cite{misner}. It implies that the energy
momentum is automatically conserved for a system coupled
consistently to the geometry of space-time. In practice, the
Bianchi identity is helpful in providing an easier approach to
study the Einstein equation. For example, we can focus on, with
the help of the Bianchi identity, independent components of the
field equations. Indeed, as shown above that the Bianchi identity
in the FRW space-time implies that : the Friedmann equation, the
$a$-equation and the $d$-equation are related by the differential
Eq. (\ref{h3}). This implies that the Friedmann equation
$G_{tt}=T_{tt}$ has a less differential order than the other
components of the Einstein equation. As a result, Friedmann
equation may serve as an useful tool in solving the differential
equations.

\section{Generalized Friedmann Equation in one-dimensional formulation}

In order to derive the non-redundant Friedmann equation from the
reduced Lagrangian, we must restore the ${\bf g}_{tt}$ dependence
of the reduced Lagrangian $L$. Indeed, the ${\bf J}^{tt}$ comes
from the variation of $L$ with respect to ${\bf g}_{tt}$, $\delta
L /\delta {\bf g}_{tt} \sim  \epsilon \phi^2 J^{tt}/2$. Hence the
most convenient way to restore the ${\bf g}_{tt}$ dependence of
the reduced Lagrangian $L$ is to introduce lapse function $b(t)$
connecting the ${\bf g}_{tt}$ metric component:
\begin{eqnarray}
ds^2 &\equiv &{\bf g}_{AB} dZ^A dZ^B \equiv g_{\mu \nu} dx^\mu dx^\nu +
f_{mn}dz^mdz^n \\
 & =&  -b(t)^2dt^2
+ {a^2}(t) \Bigl( {dr^2  \over 1 - k_1 r^2} + r^2 d^3\Omega  \Bigr
) + {d^2}(t) \Bigl( {dz^2  \over 1 - k_2 z^2} + z^2 d^D\Omega
\Bigr ) . \label{eqn:frw1} \label{GFRW}
\end{eqnarray}
This metric will be called as generalized FRW (GFRW) metric for
the $4$+$D$ anisotropic space. Once the non-redundant Friedmann
equation is derived from the variation of $L$ with respect to $b$,
one can set $b=1$ and reconstruct the $b$-independent Friedmann
equation.

The non-vanishing spin connections can be listed as follows:
\begin{eqnarray}
{\bf \Gamma}^\gamma_{\mu \nu} &=& \Gamma^\gamma_{\mu \nu}  ,\\
{\bf \Gamma}^p_{mn} &=& \Gamma^p_{mn} , \\
{\bf \Gamma}^\gamma_{mn} &=& - \partial^\gamma \beta g_{mn} ,\\
{\bf \Gamma}^p_{\mu m} &=& \partial_\mu \beta \delta^p_m .
\end{eqnarray}
Here $\partial_\mu \beta \equiv {\partial_\mu d / d}$ comes with a
non-vanishing $t$-component. In addition, we will write $I =
\partial_t \beta$ for convenience from now on. Consequently, all
non-vanishing Riemannian curvature components can be listed as
follows:
\begin{eqnarray} {\bf R}^{ti}_{\,\,\,\,tj}&=&{1\over 2}
[H\dot{B}+2B(\dot{H}+H^2)]\delta^i_j ,
\label{Rti} \\
{\bf R}^{ij}_{\,\,\,\,kl}&=& (H^2B+{k_1 \over a^2} ) (\delta^i_k \delta^j_l
-\delta^i_l
\delta^j_k) \label{Rkl} , \\
{\bf R}^{tm}_{\,\,\,\,\,\,\,\,tn}&=&{1\over 2}
[I\dot{B}+2B(\dot{I}+I^2)]\delta^m_n  ,
\label{Rtm}\\
{\bf R}^{im}_{\,\,\,\,\,\,\,\,jn}&=&HI\delta^m_n \delta^i_j ,
\label{Rim}\\
{\bf R}^{mn}_{\,\,\,\,\,\,\,\,pq}&=& (I^2B+{k_2 \over d^2} )
(\delta^m_p \delta^n_q -\delta^m_q \delta^n_p )\label{Rmn} .
\end{eqnarray}
As a result, deriving all Ricci tensors ${\bf R}^{A}_{\,\,B}$ and
scalar curvature tensors ${\bf R}$ is straightforward.

Note that there is also a $t$-dependent factor $b(t)$ in the
square-root of the metric determinant $\sqrt{{\bf g}} \sim
ba^3d^D$. Hence the variational equation of $b$ can be shown to be
\cite{kao99}
\begin{equation}
 L - H {\delta L \over \delta H} - I {\delta L
\over \delta I}  + [ H {d \over  dt}+ H( 3H +DI) -\dot{H} ]
{\delta L \over \delta \dot{H}} + [ I {d \over  dt}+ I( 3H +DI)
-\dot{I} ] {\delta L \over \delta \dot{I}}
                           =\dot{\phi}^2  \label{key}  .
\end{equation}
Note that the last term $\dot{\phi}^2$ comes from the kinetic term
of the scalar Lagrangian. Indeed, there is a kinetic coupling term
for $\phi$ via $-g^{tt} (\dot{\phi})^2/2 = b^{-2}
(\dot{\phi})^2/2$. This will bring us an additional $\dot{\phi}^2$
to the left hand side of Eq. (\ref{key}). In fact, Eq. (\ref{key})
can generalized to all fields coupled to the Einstein-Hilbert
action.

In addition, variational equations of $a$ and $d$ also give
\begin{eqnarray} \label{aeq}
&&  3L- H {\delta L \over \delta H} + (H^2- \dot{H}) {\delta L
\over \delta \dot{H}} -(2H +DI + {d \over dt}) [ -(4H +DI + {d
\over  dt}) {\delta L \over \delta \dot{H}} +{\delta L \over
\delta {H}} ] -2k_1 {\delta L \over \delta k_1} =0  ,
\\ \label{deq}
 &&  DL- I {\delta L \over \delta I} + (I^2-
\dot{I}) {\delta L \over \delta \dot{I}} - (3H +(D-1)I + {d \over
dt}) [ -(3H +(D+1)I + {d \over  dt}) {\delta L \over \delta
\dot{I}} +{\delta L \over \delta {I}} ] -2k_2 {\delta L \over
\delta k_2} =0 .
\end{eqnarray}
Note that above equations also hold in the presence of the scalar
coupling once we assume $\phi(Z)=\phi(t)$.

We will first study a simple model with a constant $d$ solution.
This is a physical solution since the internal space seems to be
small from any physical observation up to date. As shown above,
the Bianchi identity indicates that the second $a$-equation
(\ref{aeq}) is derived implicitly by the first Friedmann equation
(\ref{key}) for the constant internal space model. Therefore, we
will try to solve Eq.s (\ref{key}) and (\ref{deq}) for a complete
analysis. As a result, the Friedmann equation (\ref{key}) and the
$d(t)$ equation (\ref{deq}) become
\begin{eqnarray}
& & \bar{L} - H {\delta \bar{L} \over \delta H}  +[ H {d \over
dt}+ 3H^2  -\dot{H} ] {\delta \bar{L} \over \delta \dot{H}}
                            = \dot{\phi}^2  \label{key4} , \\
\label{deq4} &&  D\bar{L} = (3H + {d \over  dt})  [ -(3H + {d
\over dt}) {\delta \bar{L} \over \delta \dot{I}}  +{\delta \bar{L}
\over \delta {I}} ] +2k_2 {\delta \bar{L} \over \delta k_2}
\end{eqnarray}
under the $d(t)=$ constant background solution. Here a bar
notation of a variable $L$ denotes a variable evaluated at $I=0$,.
Explicitly, $\bar{L} \equiv L|_{I=0}$. In addition, we must also
check whether the solution is consistent with the constraint
$\bar{L}=0$. Our results agree with the equations shown in Ref.
\cite{3}. Details will be provided in the Appendix.

\section{Stability of an Inflationary External Space}
We need to find out whether the inflationary background de Sitter
solution $H=H_0$ and $I=0$ is a possible solution to the
Kaluza-Klein induced gravity model. As a result, a stability
analysis is also needed to find our whether this background
solution is stable or not. Furthermore, following the conventional
approach, the constant internal space solution is assumed and
should be served as a reasonable ansatz. This is because that the
internal space information appears to be minor as compared to the
$4$-space counterpart. Otherwise, we would be able to measure the
impact of the internal space physics once the internal space scale
factor $d(t)$ grows to some appreciable size. In addition, we will
also focus on the $k_1=0$ flat 4-space condition which appears to
agree with latest observations \cite{k=0}.

Therefore, we will study the existence and stability problem of
the inflationary universe for the induced Kaluza-Klein model
described above. The effective Lagrangian $\bar{L}$ for the model
$L$ under the $I=0$ condition can be shown to be
\begin{equation} \bar{L} = 3 \epsilon \phi^2  [ \dot{H} +2H^2 ]
-12 (c_1+c_2+3 c_3) [ (\dot{H} +H^2)^2 + H^4] -12 (c_2+6 c_3) [
(\dot{H} +H^2)H^2 ]+{1 \over 2} \dot{\phi}^2 -V(\phi).
\label{barLi}
\end{equation}
In addition, the variations of the $I$-equation are
\begin{eqnarray}
{\delta \over \delta I} \bar{L} &=& 6DH \left \{ {\epsilon \over
2} \phi_0^2 -c_2  \left [ \dot{H} +3H^2   \right ] -12c_3 \left [
\dot{H} +2H^2  \right ]
\right \}  \label{frd3i}  , \\
{\delta \over \delta \dot{I}} \bar{L} &=& 2D \left \{ {\epsilon
\over 2} \phi_0^2-c_2 \left [ 3\dot{H} +3H^2  \right  ] -12c_3
\left [ \dot{H} +2H^2  \right ] \right \} \label{deqdi}.
\end{eqnarray}
To summarize, there are two metric equations and one $\psi$
equation for the system:
\begin{eqnarray}
\dot{\phi}^2+ H {\delta \bar{L} \over \delta H}&= & [ H {d \over
dt}+ 3H^2 -\dot{H} ] {\delta \bar{L} \over \delta \dot{H}}
                          \label{key41} , \\
\label{deq410}  (3H + {d \over  dt}) {\delta \bar{L} \over \delta
{I}}  &=&     (3H + {d \over dt})^2 {\delta \bar{L} \over \delta
\dot{I}},
 \\
&&\bar{L}=0
\end{eqnarray}
under the flat space condition and the constant internal space
solution $I=0$. The equation $\bar{L}=0$ denotes the constraint
from the constant $\psi$ equation. Here $\bar{L} \equiv L|_{I=0}$.
The Friedmann equation reads:
\begin{equation}
{1 \over 2} \dot{\phi}^2 +V = 3 \epsilon \phi^2H^2 +6 \epsilon H
\phi \dot{\phi} +12(c_1+c_2+3c_3)(\dot{H}^2 - 2H \ddot{H} - 6H^2
\dot{H}). \label{f-eqid0}
\end{equation}
At the end of this section, we will show that the perturbation
equation of the constraint $\bar{L}=0$ is consistent with the
inflationary de Sitter solution we are interested in. More
specifically, the linear perturbation equations  $\delta \bar{L}$
can be shown to vanish automatically without generating any
further constraint on the system. Therefore, there is no need to
worry about the perturbation effect of this constraint. Hence we
can add or remove the effect of this constraint anytime. This
result also indicates that the constant $\psi$ solution, the
origin of this constraint, is a very reasonable ansatz.
Furthermore, the $\phi$-equation can be shown to be
\begin{equation} \label{eqphi}
6 \epsilon \phi (\dot{H} + 2H^2) = \ddot{\phi} +3 H \dot{\phi} +V' .
\end{equation}
Let $\phi=\phi_0+ \delta \phi$ and $H=H_0+\delta H$ with $\phi_0$
and $H_0$ some constant initial states for the inflationary de
Sitter background solution.  Here $\delta \phi$ and $\delta H$
denote small perturbations against the background solutions. We
will try to extract the linear solutions to the perturbation
effect. As a result, the leading-order Friedmann equation
(\ref{f-eqid0}) gives
\begin{equation} \label{eqfriedmann}
V_0=3 \epsilon \phi_0^2 H_0^2.
\end{equation}
Similarly, the leading-order  scalar equation ({\ref{eqphi}) gives
\begin{equation}
12\epsilon \phi_0H_0^2=V_0'.
\end{equation}
Here $V'_0 \equiv V'(\phi_0)$. Therefore, we have the constraints
combined together as:
\begin{equation} \label{constraint}
\phi_0 V'_0 =4V_0=12\epsilon \phi_0^2H_0^2 .
\end{equation}
Note that $\phi V'$ counts the effective scale-dimension of the
effective potential $V(\phi)$. For example, $\phi V'=n V$ if $V=
\phi^n$ with effective dimension $n$. Therefore,  this condition
implies that the inflationary phase exists only when the effective
dimension of the effective potential is $4$ at the inflationary
phase $\phi=\phi_0$ and $H=H_0$. Since $V =\lambda  \phi^4$
represents the scale invariant potential in $4$-dimensional space,
i.e. dim ($\lambda $) $=0$ in $4$-space, this condition will be
denoted as the scaling condition \cite{kao00}. As an example, the
spontaneously symmetry breaking potential of the form:
\begin{equation} \label{V0} V(\phi)= { \lambda \over 4}
(\phi^2-\phi^2_0)^2 +6\epsilon H_0^2 (\phi^2-\phi_0^2) + 3
\epsilon H_0^2 \phi_0^2
\end{equation}
with arbitrary coupling constant $\lambda$  can be shown to be a
good candidate satisfying the scaling condition
(\ref{constraint}).

In addition, the first order perturbation equation of $H$-equation
(\ref{f-eqid0}) and $\phi$-equation (\ref{eqphi}) can be shown to
be:
\begin{eqnarray}
&& \epsilon \phi_0 \left ( \delta \dot{\phi} -H_0 \delta \phi
\right )=4(c_1+c_2+3c_3) \left ( \delta \ddot{H} + 3H_0 \delta
\dot{H} \right )- \epsilon \phi_0^2 \delta H, \\
&& \delta \ddot{\phi} +3H_0 \delta \dot{\phi} + \left (
V''(\phi_0)- 12 \epsilon H_0^2 \right ) \delta \phi =6 \epsilon
\phi_0  \left ( \delta \dot{H} +4 H_0 \delta{H} \right ).
\end{eqnarray}
Writing $\delta H=\exp [hH_0t] \delta H_0$ and $\delta \phi=\exp
[pH_0t] \delta \phi_0$, the above perturbation equations can be
written as:
\begin{eqnarray} \label{ph}
&&  \epsilon \phi_0 \left ( p -1  \right )\delta
\phi=4(c_1+c_2+3c_3)H_0 \left ( h^2 + 3h - {\epsilon \phi_0^2
\over 4(c_1+c_2+3c_3)H_0^2} \right )\delta H, \\
&& H_0 \left ( p^2 +3p +  {V''(\phi_0)- 12 \epsilon H_0^2 \over
H_0^2} \right ) \delta \phi =6 \epsilon \phi_0  \left ( h +4
\right ) \delta{H} . \label{pp}
\end{eqnarray}
In addition to the trivial solution $\delta H=0$ and $\delta
\phi=0$, consistent solutions also exist when $h=-4$ and $p=1$. If
$h=-4$, we will have the following constraint from the right hand
side of Eq. (\ref{ph})
\begin{equation}
H_0^2 = {\epsilon \phi_0^2 \over 16(c_1+c_2 +3c_3)}. \label{h31}
\end{equation}
In addition, the existence of nontrivial solution $p=1$ implies
that
\begin{equation} \label{phic}
V''(\phi_0)= 12 \epsilon H_0^2 -4 H_0^2
\end{equation}
from the left hand side equation of Eq. (\ref{pp}). As a result,
\begin{equation}
\delta H = \exp [-4H_0t]~ \delta H_0 . \label{dh}
\end{equation}
Equivalently, the linear perturbation gives
\begin{equation}
H=H_0+ \delta H_0 ~\exp [-4H_0t]
\end{equation}
as the solution to the Hubble parameter with a small deviation
from the de Sitter background.

We will show in addition that the $I$-equation (\ref{deq410}) and
the constraint equation $\bar{L}=0$ are both consistent with above
constraints in the presence of static internal space and de Sitter
external space. Indeed, writing $D_t \equiv \partial_t +3H$,
$\delta \bar{L} / \delta {I}\equiv \bar{L}_I$ and $\delta \bar{L}
/ \delta \dot{I} \equiv \bar{L}_{\dot{I}}$, the $I$-equation
$$
(3H + {d \over  dt}) {\delta \bar{L} \over \delta {I}}  = (3H + {d
\over dt})^2 {\delta \bar{L} \over \delta \dot{I}},$$ under the
condition of flat and static internal space can be integrated to
give
\begin{equation}
Y(H) \equiv \bar{L}_{{I}}-D_t\bar{L}_{\dot{I}}=6 D(c_2+4c_3)
[\ddot{H} +4H \dot{H} ] + { K_1 \over a^3}=0
\end{equation}
with a constant of integration $K_1$. During the inflationary
phase, we can ignore the effect of $K_1$ term. Therefore, the
$I$-equation does not have a leading order contribution. In
addition, the first order perturbation of this equation gives
$\delta Y=6D(c_2+4c_3) [\delta \ddot{H} +4H_0 \delta \dot{H} ]$
which vanishes identically in compatible with the perturbation
$\delta H$ given in Eq. (\ref{dh}). In fact, we can also compute
the complete $I$-equation and find that
\begin{equation}
D_t Y =6D (c_2+4c_3) [\dot{\ddot{H}}+ 4 {\dot H}^2+ 7H \ddot{H}
+12H^2 \dot{H} ]=0.
\end{equation}
There is no leading contribution either. In addition, the
perturbative equation takes the form
\begin{equation}
\delta D_t Y=D_t \delta Y= 6D(c_2+4c_3) D_t[\delta \ddot{H} +4H_0
\delta \dot{H} ]=6D (c_2+4c_3) [\delta \dot{\ddot{H}}+ 7H \delta
\ddot{H} +12H^2 \delta \dot{H} ]=0,
\end{equation}
that vanishes identically with $\delta H$ given in Eq. (\ref{dh}).

As mentioned above, we still have to compute all possible
constraints from the internal space scalar field $\psi$-equation
$\bar{L}=0$. It is interesting to find that the leading order
perturbation equation for $\bar{L}=0$ gives
\begin{eqnarray}
6 \epsilon \phi_0^2 H_0^2 - V_0 =3 \epsilon \phi_0^2 H_0^4= 12
(2c_1+3c_2+12c_3) H_0^4 \label{barL1}
\end{eqnarray}
incorporating the scaling constraint $V_0= 3 \epsilon
\phi_0^2H_0^2$. Therefore, the leading order equation of
$\bar{L}=0$ gives another constraint
\begin{equation}
H_0^2 = {\epsilon \phi_0^2 \over 4(2c_1+3c_2+12c_3)}. \label{hL}
\end{equation}
In addition, the first order perturbation of equation this $\psi$
constraint can be shown to be
\begin{equation}
\left [  3 \epsilon \phi_0^2 -12 (2c_1+3c_2+12 c_3) H_0^2   \right
] \left [ \delta \dot{H} +4H_0 \delta H \right ] + (12\epsilon
\phi_0H_0^2 -V_0') \delta \phi=0 . \label{barLh}
\end{equation}
Therefore, this equation is completely consistent with the whole
system by observing that all coefficients in above equation
vanishes identically. Indeed, $3 \epsilon \phi_0^2 -12
(2c_1+3c_2+12 c_3) H_0^2 =0$ following Eq. (\ref{hL}). In
addition, $\delta \dot{H} +4H_0 \delta H =0$ and $12\epsilon
\phi_0H_0^2 -V_0'$ are automatically satisfied.

In summary, the constraints (\ref{h31}) and (\ref{hL}) indicate
that the coupling constants should obey the following constrain in
order to admit an inflationary phase in the presence of a static
internal space:
\begin{equation}
2c_1+c_2=0.
\end{equation}
As a result, the Hubble constant and the field parameters are
related by:
\begin{equation}
H_0^2 = {\epsilon \phi_0^2 \over 8(c_2+6c_3)}. \label{hall}
\end{equation}
Note that above results indicate that: (1) static internal flat
space solution is completely compatible with the conditions of
inflationary solution;  (2) the perturbation equation of
$\bar{L}=0$ is a perfect identity consistent with all other
constraints derived elsewhere. This indicates that the constant
$\psi$ solution is a very reasonable choice for the stationary
state of the system. In addition, (3) the static internal space
assumptions is also a consistent choice for the existence and
stability of the inflationary phase.

In addition, we can also consider the perturbation of $\psi$ by
setting $\psi=\psi_0+\delta \psi$ and perturb the field equation
(\ref{psieq}). The result is $ D_A \partial^A  \delta \psi =0.$
Here we have also used the identity $\delta L=0$ under the linear
perturbation shown above. Assuming $\delta \psi(Z)=\delta\psi (z)$
such that the internal space $z$ is completely decoupled from the
$4$-D space time. As a result the perturbation equation $ D_A
\partial^A  \delta \psi =0$ is consistent if $\delta \psi(z)$ is
a harmonic function obeying $\partial^m\partial_m \delta \psi=0$.
Hence $\delta \psi(z) =$ constant for a consistent perturbation.
Therefore, the constant $\psi$ is also a consistent choice of
background solution.

Note also that the effect of the $\psi=$constant implies that
$L=0$. Together with the constraint $2c_1+c_2=0$ for the existence
of an inflationary phase, one has effectively a scalar equation of
the form
\begin{equation}
 {1 \over 2}
\partial_A \phi \partial^A \phi + V(\phi) +{\epsilon \over 2}
\phi^2 {\bf R} =   {c_2 \over 2} ({\bf R}^{AB}_{\;\;\;\;\;\;CE})^2
- c_2 ({\bf R}^{A}_{\;\;\;B})^2 - c_3 {\bf R}^2 \label{pdg1}
\end{equation}
with a purely geometric source. Although the equation $L=0$ is not
exactly a Klein-Gordon equation of the form $(\partial^2 +m^2 +
R/5) \phi =\alpha R^2$ studied in Ref. \cite{k1}, both theories
appear to have similar physical origin. It was shown that an
effective re-normalized Lagrangian of the form $\alpha \phi R^2$
is a result of dimensional consideration. Indeed, the coupling
constant $\alpha$ can only be made dimensionless, rendering a
system free from introducing any additional arbitrary length
scale, if the space-time dimension $N=6$ \cite{k1}.

Note that the scalar fields $\psi$ and $\phi$, both with dimension
one, considered in this paper are designed to replace all
dimensionful coupling constants with appropriate scalar fields. As
a result, all coupling constants are assumed to be dimensionless
in this approach. The only exception is some parameters associated
with the SSB potential $V(\phi)$ designed to pick up a symmetry
breaking scale. The constraint equation $L=0$ indicates that the
scalar field $\psi$ introduced here may have close relation with
the re-normalizability of the energy-momentum tensor for $\phi$.
In addition, similar model has been studied in Ref. \cite{k2} with
Gauss-Bonnet (GB) Lagrangian coupled to a perfect fluid. The
constraint $2c_1+c_2=0$ in this paper follow first from the
stability of the Friedmann equation (\ref{f-eqid0}) with a
coefficient of the combination $(c_1+c_2+3c_3)$ coupled to the
quadratic interactions $\dot{H}^2 - 2H \ddot{H} - 6H^2 \dot{H}$.
This coefficient vanishes for GB term with $c_1:c_2:c_3=1:-4:1$.
This is the main difference between the model considered in Ref.
\cite{k2} and current models. In addition, we have also shown in
the Appendix that the Friedmann equation agrees with Ref. \cite{3}
up to a difference in the definition of sign of the coupling
constants $c_i$. Our inflationary phase solution also agrees with
Ref. \cite{3} up to a scale due to the effect of the scalar field
in this induced gravity model. Indeed, the Friedmann equation and
the $I$-equation implies that $H_0^2=\Lambda/6$ and
$(c_2+4c_3)\Lambda=1$ respectively after setting $V=\Lambda$ for a
system without a dynamical scalar field $\phi$. As a result,
$H_0^2=1/[6((c_2+4c_3)]$.

\section{Conclusion}

It is shown that replacing the internal space dimensionful
coupling constant with a dimension one scalar field $\psi=$
constant works harmonically with the Kaluza-Klein inflationary
universe under the constraint $2c_1+c_2=0$. In addition, from the
effective Lagrangian shown in Eq. (\ref{barLi}), it is easy to
find that any quadratic Lagrangian must present itself as
combinations of the form: $\bar{L}_2= l_1\dot{H}^2+
l_2(\dot{H}H^2+H^4) \equiv -12 (c_1+c_2+3 c_3) [ (\dot{H} +H^2)^2
+ H^4] -12 (c_2+6 c_3) [ (\dot{H} +H^2)H^2 ]$ with dimensionless
$l_i$ corresponding linear combinations of $c_i$ defined
accordingly. Here $\bar{L}_2$ denotes the quadratic part of the
effective Lagrangian $\bar{L}$. As a result, it can be shown that
any quadratic Lagrangian of the combinations ${\bar L}_2 \sim
l_1\dot{H}^2+ l_2(\dot{H}H^2+H^4)$ will not contribute to the
Friedmann equation.

Indeed, the quadratic terms contribute to the Friedmann equation
(\ref{key41}) according to
\begin{eqnarray} \label{Lg}
E_2=\bar{L}_2 +H ( {d \over dt} +3H )\bar{L}_{\dot{H}} -
H\bar{L}_{H} - \dot{H} \bar{L}_{\dot{H}} \to \bar{L}_2 +3H^2
\bar{L}_{\dot{H}} - H\bar{L}_H
\end{eqnarray}
in the de Sitter background with $L_H \equiv \delta
\bar{L}_2/\delta H$ and $L_{\dot{H}} \equiv \delta
\bar{L}_2/\delta \dot{H}$ shown in above equation.  It is clear
that the $l_1$ term does not contribute to above equation $E_2$ in
the de Sitter space with $\dot{H}_0=0$. Therefore, we have
effectively the quadratic Lagrangian $\bar{L}_2 =
l_2(\dot{H}H^2+H^4)$ needed to be considered for its effect on the
leading order Friedmann equation. As a result, we can show that
$\bar{L}_2 \to  l_2H_0^4$, $H^2 \bar{L}_{\dot{H}} \to l_2 H_0^4$
and $H\bar{L}_H \to 4l_2H_0^4$. Hence the total contribution of
the quadratic Lagrangian to $E_2$ cancels each other. Therefore,
this proves that the quadratic Lagrangian does not contribute to
the de Sitter solution in four dimension.

The perturbation equation for $\delta \phi$ indicates a constraint
(\ref{phic}) $ V''(\phi_0)= 12 \epsilon H_0^2 -4 H_0^2$ which
turns out to be inconsistent with the SSB scalar potential
(\ref{V0}). Indeed, $V''_0=2 \lambda \phi_0^2+12 \epsilon H_0^2$
for this potential. Hence the constraint (\ref{phic}) implies that
$\lambda \phi_0^2=-2H_0^2$. A negative $\lambda$ indicates that
the SSB potential is an unstable potential without a global
minimum. In fact, we can show that the local minimum is at
$\phi_m^2=0$ and the local maximum is at $\phi^2_M=(1+6
\epsilon)\phi_0^2$. Hence the consistent initial state $\phi_0$ of
the scalar field will be expected to locate at the left hand side
of the maximum point $\phi_M$. The scalar field will hence roll
down to the local minimum which is located at $\phi=0$. This will
lead to an un-physical state with infinite Newtonian constant $G$.
In addition, this local minimum is also not a stable vacuum state.
$\phi$ will eventually tunnel to its global minimum at $\phi \to
\infty$. Therefore, the constraint (\ref{phic}) is not a physical
constraint for the SSB potential and this is also true for the
Coleman-Weinberg effective potential \cite{coleman}. Hence the
only consistent perturbation of $\phi$ is $\delta \phi=0$. This
indicates that the de Sitter background is highly stable and
compatible with the stable mode $\delta H \to 0$. Therefore, the
system will remain stable as long as the scalar field does change
very slowly. Note that the negative coupling constant appears to
be a universal features of any coupled effective potential
including the Eric-Weinberg dynamical symmetry breaking potential
\cite{weinberg}.

Indeed, the $\phi$ equation under the slow rollover assumptions,
$|\dot{\phi}/\phi|\ll H_0$ and $|\ddot{\phi}/\phi| \ll H_0^2$,
states that
$$\ddot{\phi} +3 H_0 \dot{\phi} \sim 0 $$
during the period where $H \sim H_0$. This gives
\begin{equation}
\phi \sim \phi_0 + {\dot{\phi}_0 \over 3 H_0}[1 - \exp (- 3 H_0 t)
]. \label{phit}
\end{equation}
Therefore, the slow rollover-approximation is indeed consistent
with the dynamics of the scalar field equation that has been a
focus of research interests in the literature. Therefore, $\phi$
does change very slowly during this inflationary phase.

In summary, we have derived abundant constraints from the
assumptions: (i) $\psi(z)=$ constant, (ii) $d(t)=$ constant. These
assumptions are adopted partly from the fact that they are both
not appreciable in the four dimensional physical universe observed
today. Hence it is reasonable to freeze their dynamics at certain
stage of the evolutionary process. The abundant consistency shown
in this paper compatible with the abundant constraints from these
assumptions implies that these assumptions are in fact rather
reliable. Although we are unable to provide a dynamical reason for
these assumptions from the first principle, compatibility of these
assumptions with the inflationary four space deserves more
attention for its possible physical implications.

Our result indicates, however, that both assumptions, static
$\psi$ and static $d$, appear to be a consistent set of choice for
the higher derivative Kaluza-Klein models. As a result, the
Kaluza-Klein higher derivative induced gravity theory behaves
similar to the conventional four dimensional induced higher
derivative gravity in the lower-energy limit namely, only $R^2$
and $R_{ab}^2$ couplings remain effective during the inflationary
de Sitter phase. This is in consistent with the $4$-D theories
that $R_{abcd}^2$ terms can be replaced by $R^2$ and $R_{ab}^2$
couplings following the GB theorem. Therefore, related research
deserves more attention.

\section*{Acknowledgments}
This work is supported in part by the National Science Council of
Taiwan.

\appendix
\section{Appendix}
\subsection{Field Equation}
The field equation of the Lagrangian  $L= - {\epsilon} \phi^2 {\bf
R} /2  -c_1 ({\bf R}^{AB}_{\;\;\;\;\;\;CE})^2 - c_2 ({\bf
R}^{A}_{\;\;\;B})^2 - c_3 {\bf R}^2-\partial_A \phi
\partial^A \phi/2 -V$ can be derived by the variation of ${\bf g}_{AB}$. The
result is:
\begin{eqnarray}
 & &  {\epsilon \phi^2 \over 2} \left ( {1 \over 2}
{\bf R}{\bf g}_{AB} - {\bf R}_{AB} \right ) +{1\over 2}{\bf
g}_{AB} \left ( c_1 ({\bf R}^{AB}_{\;\;\;\;\;\;CE})^2 + c_2 ({\bf
R}^{A}_{\;\;\;B})^2 +c_3 {\bf R}^2 -L_\phi \right )
\nonumber \\
&=&  2(c_3 {\bf R}\,{\bf R}_{AB}+c_2 {\bf R}_{AC}{\bf R}^C_B+c_1
{\bf R}_{ACDE}{\bf R}_B^{CDE} ) -
2c_3 ({\bf g}_{AB}D^2 - D_AD_B){\bf R} -{c_2 \over 2}{\bf g}_{AB}D^2 {\bf R} \nonumber \\
&&- (4c_1+c_2)  D^2 {\bf R}_{AB} + 2(2c_1+c_2) D_AD_C{\bf
R}_B^{\;C} + {\epsilon \over 2}  ( D_A \partial_B- {\bf
g}_{AB}D^2) \phi^2 +{1 \over 2} \partial_A \phi \partial_B \phi.
\label{eom}
\end{eqnarray}
In addition, the scalar equation can be shown to be:
\begin{equation}
D^2 \phi =V'+ \epsilon \phi {\bf R} \label{phi}.
\end{equation}
In order to derive the field equation in a covariant way, we may
write the variation of the Riemann curvature tensor as $\delta
{\bf R}^D_{\;\; CBA} =
-D_A\delta\Gamma^D_{BC}+D_B\delta\Gamma^D_{AC}$ as if
$\delta\Gamma^A_{BC}$ is a type T(1,2) tensor. The derivation has
nothing to do with whether $\delta\Gamma^A_{BC}$ is a tensor or
not. Rather, by imagining $\delta\Gamma^A_{BC}$ is a tensor and
use all related properties of tensor, it helps reducing the effort
in deriving these equations, especially when integration-by-part
is required. In addition, we have also used the Bianchi identity
$D_CD_D{\bf R}^{ACDB}= D^2 {\bf R}^{AB}- D_CD^A{\bf R}^{BC}$ in
converting the differentiation of Riemann tensor into
differentiation of Ricci tensor. Note also that our result agrees
with Ref. \cite{3} up to a difference in definition of sign in
$c_i$. In particular, we can show explicitly that the Friedmann
equation for static internal space agrees with Eq. (15) in Ref.
\cite{3}.

In addition, the static $I$-equation (\ref{deq4}) can be written
as
\begin{equation}
D \bar{L}=D_t Y =6D (c_2+4c_3) D_t [\ddot H + 4 H \dot H ]=6D
(c_2+4c_3) D_t \partial_t [\dot H + 2 H^2 ]= -D (c_2+4c_3)D^2 R.
\end{equation}
Here $D_t \equiv \partial_t +3H$. As a result, the static internal
space $I$-equation can be written as
\begin{equation}
\bar{L}+ (c_2+4c_3)D^2 R=0 \label{eqImn}
\end{equation}
which is identical with the ${mn}$ component of the Einstein
equation (\ref{eom}). Note that $R_{mn}=R_{an}=0$ in the static
internal space condition. Therefore, the only thing left over from
the $mn$ component Einstein equation is identical to Eq.
(\ref{eqImn}). Note that this result agrees with the Eq. (17) in
Ref. \cite{3}.

As shown earlier, ${\bf J}_{AB}={\bf G}_{AB}-{\bf T}_{AB}={1 \over
2} {\bf R}{\bf g}_{AB} - {\bf R}_{AB}-{\bf T}_{AB}=0$ can be
decomposed into three different equations: ${\bf J}_{tt}=0$, ${\bf
g}^{ij}{\bf J}_{ij}=0$ and ${\bf g}^{mn}{\bf J}_{mn}=0$ in the
$4+D$ dimensional space-time described by the GFRW metric
(\ref{FRW}). These equations correspond to the Friedmann equation
(\ref{key}), $a$-variational equation (\ref{aeq}), and the
$d$-variational equation (\ref{deq}). Therefore, this proves that
Eq.s (\ref{key}-\ref{deq}) and (\ref{eqphi}) are the complete set
of field equations in the GFRW space-time. We choose to ignore one
of the redundant equation (\ref{aeq}) in this paper without
loosing any physical information as implied by the Bianchi
identity (\ref{h3}).

In summary, the reduced formulae shown in this paper can be
helpful in extracting some useful information without going into
the details of the field equations. For example, the existence of
the inflationary solution $H=H_0$ has to do with the leading order
equations. It can be done by ignoring any term like $f(H)\dot{H}$,
with $f(H)$ arbitrary function of $H$. On the other hand, the
stability of the inflationary solution has to do with those
leading order terms linear in time differentiation of $\delta
{H}$. We can freely ignore terms like $\dot{H}^2$. In particular,
$(d/dt)(f(H)\delta H)= f(H) \delta \dot{H}$ can be used to skip
unrelated terms, with $f(H)$ arbitrary function of $H$, with the
close formula shown in this paper.

\subsection{Curvature Tensor}
For completeness of the calculation, we will list all
non-vanishing components of the Ricci tensors, scalar curvatures
and terms present in the Lagrangian:
\begin{eqnarray}
{\bf R}^{ti}_{\,\,\,\,tj}&=& [\dot{H}+H^2]\delta^i_j ,
\label{Rti} \\
{\bf R}^{ij}_{\,\,\,\,kl}&=& (H^2+{k_1 \over a^2} ) (\delta^i_k
\delta^j_l -\delta^i_l
\delta^j_k) \label{Rkl} , \\
{\bf R}^{tm}_{\,\,\,\,tn}&=& (\dot{I}+I^2)\delta^m_n  ,
\label{Rtm}\\
{\bf R}^{im}_{\,\,\,\,jn}&=&HI\delta^m_n \delta^i_j ,
\label{Rim}\\
{\bf R}^{mn}_{\,\,\,\,pq}&=& (I^2+{k_2 \over d^2} ) (\delta^m_p
\delta^n_q -\delta^m_q \delta^n_p )\label{Rmn} .
\end{eqnarray}
In addition, one has:
\begin{eqnarray}
{\bf R}^{t}_{\,\,t}&=& -[3  (\dot{H}+H^2) +D(\dot{I}+I^2)] ,
\label{Rtt}\\
{\bf R}^{i}_{\,\,j}&=&- [\dot{H}+3H^2+2{k_1 \over a^2}+
DHI]\delta^i_j
\label{Rij}, \\
{\bf R}^{m}_{\,\,n}&=& -[\dot{I}+DI^2+(D-1){k_2 \over d^2}+3HI]
\delta^m_n ,
\label{Rm} \\
{\bf R}&=& - [ 6(\dot{H}+2H^2+{k_1 \over a^2}+
DHI)+D(D-1)(I^2+{k_2 \over d^2}) +2D(\dot{I}+I^2)], \label{R}
\end{eqnarray}
and
\begin{eqnarray}
\label{R42} ({\bf R}^{AB}_{\;\;CD})^2 &=& 12 (\dot{H} +H^2)^2 +4D
(\dot{I} +I^2)^2 + 12DH^2I^2
+12 (H^2 + {k_1 \over a^2})^2 +2D(D-1)(I^2 + {k_2 \over d^2})^2  ,\\
\label{R22} ({\bf R}^A_{\;B})^2 &=& 12 (\dot{H} +H^2)^2 +D(D+1)
(\dot{I} +I^2)^2 +12 (H^2 + {k_1 \over a^2})^2 +D(D-1)^2(I^2 +
{k_2 \over d^2})^2 + 3D(D+3)H^2I^2
\nonumber \\
&+& 12 (\dot{H} +H^2) (H^2 + {k_1 \over a^2}) +6D(\dot{H}+H^2
+HI)(\dot{I}+I^2)
+6DHI (\dot{H}+3H^2+2{k_1 \over a^2}) \nonumber \\
&+& 2D(D-1)(\dot{I}+ I^2)(I^2 + {k_2 \over d^2}) +6D(D-1)HI(I^2 +
{k_2 \over d^2}) ,
\\
\label{R02} ({\bf R})^2 &=& 36 (\dot{H} +H^2)^2 +4D^2 (\dot{I}
+I^2)^2 + 36D^2H^2I^2 +36 (H^2 + {k_1 \over a^2})^2
+D^2(D-1)^2(I^2 + {k_2 \over d^2})^2
\nonumber \\
&+& 72 (\dot{H} +H^2) (H^2 + {k_1 \over a^2}+ DHI)
+12D(D-1)(\dot{H}+2H^2+ {k_1 \over a^2} )(I^2 + {k_2 \over d^2})
\nonumber \\
&+&24D(\dot{H}+2H^2+{k_1 \over a^2})(\dot{I}+I^2) +72DHI (H^2+{k_1
\over a^2}) +4D^2(D-1)(\dot{I}+ I^2)(I^2 + {k_2 \over d^2})
\nonumber \\
&+& 24D^2(\dot{I}+ I^2)HI +12D^2(D-1)HI(I^2 + {k_2 \over d^2}) .
\end{eqnarray}

\subsection{Compactification}
Another way to derive the decoupled field equation is to assume
that the $N$-space decouples according to $M^N \to M^4 \times M^D$
via the following metric decomposition:
\begin{equation}
ds^2 = g_{ab}(x) dx^a dx^b + d^2(x) h_{mn}(z)dz^mdz^n
\end{equation}
with internal space metric $g_{mn}(x,z)=d^2(x) h_{mn}(z)$.

Non-vanishing spin connections are:
\begin{equation}
{\bf \Gamma}^a_{bc}= {\Gamma}^a_{bc}; {\bf \Gamma}^c_{mn}=-E^c
g_{mn}; {\bf \Gamma}^m_{cn}=E_c \delta^m_n; {\bf
\Gamma}^m_{np}=\Gamma^m_{np},
\end{equation}
with $E_c \equiv \partial_cd(x)/d(x)$ a vector like function. As a
result, we can also show that all non-vanishing curvature tensors
are:
\begin{eqnarray}
{\bf R}^{ab}_{\;\;\;\; cd}&=& R^{ab}_{\;\;\;\; cd} ,\\
{\bf R}^{am}_{\;\;\;\; bn}&=& -\nabla^aE_b \delta^m_n , \\
{\bf R}^{mn}_{\;\;\;\; pq}&=& - E^aE_a  (\delta^m_p
\delta^n_q -\delta^m_q \delta^n_p )\label{RmnN}, \\
{\bf R}^a_{\;\;b} &=& R^a_{\;\;b} + D \nabla^aE_b, \\
{\bf R}^m_{\;\;n} &=& (D_a+DE_a)E^b \delta^m_n,  \\
{\bf R} &=& R + 2D \nabla'_a  E^a
\end{eqnarray}
with $\nabla_aE_b \equiv (D_a+E_a)E_b$ and $\nabla'_aE_b \equiv
[D_a+(D+1)/2\;E_a]E_b$.

The Lagrangian $L= - {\epsilon \over 2} \phi^2 {\bf R}  -c_1 ({\bf
R}^{AB}_{\;\;\;\;\;\;CE})^2 - c_2 ({\bf R}^{A}_{\;\;\;B})^2 - c_3
{\bf R}^2-\partial_A \phi
\partial^A \phi/2 -V$ can be shown to be:
\begin{eqnarray}
 & &  L(E_a) ={\epsilon \phi^2 \over 2} \left[  R + 2D
\nabla'_a E^a \right ] - c_3 \left [ R^2+ 4D R \nabla'_a E^a +4D^2
\left ( \nabla'_a E^a \right )^2 \right ] -\partial_A \phi
\partial^A \phi/2 -V\\
&& - c_2 \left (
{R^{a}_{\;\;\;b}} ^2 + 2DR^{ab} \nabla_aE_b +D^2 [(D_a+DE_a)E_b ][(D^a+DE^a)E^b ] \right ) \\
&& -c_1 \left [  { R^{ab}_{\;\;\;\;cd}}^2 +4 D (\nabla_aE^a)^2
+2D(D-1) (E_aE^a)^2  \right ]\label{eomN} .
\end{eqnarray}
Note that $E_a=\partial_a d/d$, therefore the $I$-equation can be
derived by varying above effective Lagrangian with respect to $d$.
We can first derive the field equation with respect to $\delta
E_a$ and perform another integration-by-part to find the field
equation of $\delta d$.  Note also that only terms linear in $E_a$
will contribute to the $I$-equation once $I=0$ is imposed for
static internal space solution. Therefore, we only need to
consider
\begin{eqnarray}
 & &  L=   D \phi^2
D_a E^a - 4c_3 D R D_a E^a   - 2c_2
 DR^{ab} D_aE_b \sim D \left [
 (c_2+4c_3) D_aR -  D_a \phi^2\right ]  E^a \label{eomNI} .
\end{eqnarray}
In addition, there is also a term derived from the volume measure
$\sqrt{\bf g} \propto d^D$. Therefore, the $I$-equation can be
shown to be:
\begin{equation}
D \left [ \bar{L}- (c_2+4c_3)D^2 R \right ] =0 \label{eqIN} .
\end{equation}
This is identical to the $I$-equation (\ref{eqImn}) $\bar{L}+
(c_2+4c_3)D_t\partial_t R=0$ shown above. And also note that Eq.
(\ref{eqIN}) holds for the case with $d=d(x)$ in the presence of a
constant background internal space $d=$ constant. Therefore, the
conditions of inflationary phase remain the same for inhomogeneous
$d(x)$ in the presence of a static internal background.


\end{document}